\documentclass{article}

\usepackage[final, nonatbib]{neurips_2023_ml4ps}
\usepackage[numbers]{natbib}

\usepackage[utf8]{inputenc} 
\usepackage[T1]{fontenc}    
\usepackage{hyperref}       
\usepackage{url}            
\usepackage{booktabs}       
\usepackage{amsfonts}       
\usepackage{nicefrac}       
\usepackage{microtype}      
\usepackage{xcolor}     
\usepackage{graphicx} 

\title{Domain Adaptation for Measurements of Strong Gravitational Lenses}

\author{
  Paxson ~Swierc$^1$\\
  Department of Astronomy and Astrophysics\\
  University of Chicago\\
  \texttt{pswierc@uchicago.edu} \\
   \And
  Megan ~Zhao$^1$ \\
  Department of Astronomy and Astrophysics\\
  University of Chicago\\
  \texttt{yifanzyf@uchicago.edu} \\
   \And
  Aleksandra ~Ćiprijanović \\
  Computational Science and AI Directorate\\ 
  Fermi National Accelerator Laboratory;\\ 
  Department of Astronomy and Astrophysics\\
  University of Chicago\\ 
  \texttt{aleksand@fnal.gov} \\
   \And
   Brian Nord\\
  Computational Science and AI Directorate\\
    Fermi National Accelerator Laboratory;\\
    Department of Astronomy and Astrophysics\\
    University of Chicago;\\
    Kavli Institute for Cosmological Physics\\ University of Chicago\\
    \texttt{nord@fnal.gov}
}

\begin{document}

\maketitle

\begin{abstract}
  Upcoming surveys are predicted to discover galaxy-scale strong lenses on the order of 10\textsuperscript{5}, making deep learning methods necessary in lensing data analysis. 
  Currently, there is insufficient real lensing data to train deep learning algorithms, but the alternative of training only on simulated data results in poor performance on real data. 
  Domain Adaptation may be able to bridge the gap between simulated and real datasets. 
  We utilize domain adaptation for the estimation of Einstein radius ($\Theta_E$) in simulated galaxy-scale gravitational lensing images with different levels of observational realism. 
  We evaluate two domain adaptation techniques -  Domain Adversarial Neural Networks (DANN) and Maximum Mean Discrepancy (MMD). 
  We train on a source domain of simulated lenses and apply it to a target domain of lenses simulated to emulate noise conditions in the Dark Energy Survey (DES). 
  We show that both domain adaptation techniques can significantly improve the model performance on the more complex target domain dataset. 
  This work is the first application of domain adaptation for a regression task in strong lensing imaging analysis.
  Our results show the potential of using domain adaptation to perform analysis of future survey data with a deep neural network trained on simulated data. 

\end{abstract}

\section{Introduction}

\footnotetext[1]{Equal contribution.}
Strong gravitational lensing is a powerful probe of astrophysics, as well as cosmology. 
Galaxy-scale strong lensing systems in the Sloan Lens ACS (SLACS)~\cite{auger2010sloan} survey gave insights into the dark and baryonic matter profiles in massive elliptical galaxies~\cite{auger2010sloan, barnabe2011two}. 
In cosmology, strong lensing systems have been used as an independent probe of the Hubble constant through time delays in multiple images in the system~\cite{Birrer_2022, suyu2010dissecting} and for constraining the   
dark energy equation of state~\cite{collett2014cosmological, sereno2002probing}. 

Upcoming large-scale astronomical surveys, including the Rubin Observatory's Legacy Survey of Space and Time (LSST)\cite{ivezic2019lsst}, the Hyper Suprime-Cam Subaru Strategic Program (HSC)\cite{Aihara_2017}, Euclid\cite{2019clrp.2020...20P}, and the Nancy Grace Roman Space Telescope~\footnotemark[2] \footnotetext[2]{https://www.jpl.nasa.gov/missions/the-nancy-grace-roman-space-telescope}, will observe an unprecedented number of lensed systems. 
Collett~\cite{Collett_2015} predicted that $>10^5$ galaxy scale strong lensing systems will be observed by LSST and Euclid. 
The large amount of data from these surveys will require much more efficient analysis techniques, such as those utilizing machine learning (ML) techniques. 
Currently discovered lensing systems are on the order of a thousand, which is inadequate to train many ML algorithms, forcing researchers to use simulated datasets of various forms, such as in~\cite{zaborowski2022identification, Rezaei_2022, Wilde_2022}. 
For classification tasks, it has been shown that the inevitable gap between training and testing data causes complex, large-parameter ML models trained on labeled simulation data to perform poorly (i.e., systematic biases) on new, unlabeled observational data \cite[e.g.,][]{avestruz2019automated}.
The same issues with classification are present in regression tasks, requiring an advancement to mitigate biases in deep learning analyses.


Domain adaptation (DA) is a class of techniques used when the distribution of training data differs from that of the testing data~\cite{CS2017}. 
In DA, the training distribution of data is the source data, while the data it will be tested on is the target data. 
We are specifically evaluating unsupervised domain adaptation~\cite{SN2023}, which is the problem scenario where one has access to labeled source data, but target data is unlabeled. 
Unsupervised DA techniques adjust training to include unlabeled target data, detect the difference between the source and the target domain distributions, and aim to minimize it.
In astronomy, DA has been applied to multiple scenarios of classification ~\cite[e.g.,][]{ciprijanovic2021deepmerge,CK2022,CL2023,alexander2021domain,Vilalta_2019}), generative models ~\cite[e.g.,][]{Lin_2021, O_Briain_2021}, and regression in one scenario~\cite{GD2021}. 

We apply two DA techniques to strong gravitational lensing analysis in this work: adversarial training using Domain Adversarial Neural Networks (DANNs)~\cite{ganin2016domain} and a distance-based method called Maximum Mean Discrepancy (MMD)~\cite{JMLR:v13:gretton12a}. 
We apply both techniques to the task of estimating the Einstein radius ($\Theta_E$) of galaxy-scale gravitational lensing systems. 
Both source and target domain data are simulated using {\tt deeplenstronomy}\footnote{https://github.com/deepskies/deeplenstronomy}~\cite{morgan2021deeplenstronomy}. 
The source data is simulated strong lenses, with no noise. 
For target data, the same strong lenses are simulated, but with emulated Dark Energy Survey (DES)\cite{PhysRevD.98.043526} conditions. 
We show that both techniques improve the model performance on the target domain, demonstrating the capacity of using DA to bridge the gap between simulated and real data.

Our manuscript is organized as follows. In Section~\ref{sec:methods} we describe the two DA methods utilized in this work; in Section~\ref{sec:data} we describe the data; in Section~\ref{sec:NN Architecture} we describe the neural network model; in Section~\ref{sec:results} we present our results; and in Section~\ref{sec:discussion} we discuss our findings.

\section{Methods}
\label{sec:methods}

Both DA approaches used are unsupervised techniques: while they require labeled data from the source domain, they leverage only unlabeled data from the target domain. 
This mirrors the problem setting of training on labeled simulated data, while only having access to unlabeled real data. 
We assume that the two domains have the same conditional probability distributions of a label $y$, given some data $x$, ${p}_{s}{(y|x)} = {p}_{t}{(y|x)}$, but that the marginal probability distribution of the data $x$ within the domains themselves is not equal: ${p}_{s}{(x)} \neq {p}_{t}{(x)}$. 
This is known as a covariate shift between domains~\cite{farahani2021brief}. 
Under this assumption, feature-based DA techniques such as DANN and MMD aim to close the gap by learning a transformation that maps source data into target data~\cite{farahani2021brief}. 
DANN and MMD find this transformation by utilizing domain invariant features. 
In practice, a Convolutional Neural Network (CNN) feature extractor learns to produce invariant features that are identical between the two domains.

We use a total loss:
$\mathcal{L}_{Total}=\mathcal{L}_{MSE} + \lambda_{DA}\mathcal{L}_{DA}$, where $\mathcal{L}_{MSE}$ is the Mean Squared Error (MSE) loss used to minimize the error on the task the model is trained to perform (estimate the Einstein radius of a gravitational lens). 
$\mathcal{L}_{DA}$ is either the MMD loss or the DANN loss, and it minimizes the variance of extracted features between domains. 
This loss is weighted by constant $\lambda_{DA}$. 

\textbf{DANN}~\cite{ganin2016domain} uses a two-headed network, with an adversarial approach to find invariant features. 
All data is fed into a feature extractor made of convolutional layers, which is then forwarded to the two heads, both made of linearly connected layers. 
The first head is the label predictor, which takes features and makes predictions of the labels associated with them. 
This head is only trained with features extracted from labeled source data and minimizes $\mathcal{L}_{MSE}$. 
The second head is the domain classifier, which takes features from both domains and makes predictions as to which domain those features came from. We use a Negative Log-Likelihood Loss function (NLL) loss as $\mathcal{L}_{DA}$. 
With DANNs, the goal is to learn domain-invariant features, which will maximize $\mathcal{L}_{DA}$ loss and confuse the domain classifier.
To simultaneously optimize the 
the feature extractor to find features that will enable classification of labeled data in the label predictor head, and confuse the domain classifier head (rendering it unable to differentiate between the domains), a gradient reversal layer is used in between the feature extractor and the domain classifier. 
This layer does not affect the data in forward propagation, but multiplies the gradient by -1 in back propagation, flipping the gradient.


\textbf{MMD}~\cite{JMLR:v13:gretton12a} is a distance-based domain adaptation method that identifies invariant features by minimizing the distance between the source and target domain distributions. 
We employ kernel methods to calculate MMD distance, which we use as our DA loss:

$$\mathcal{L}_{DA}=\frac{1}{N(N-1)} \sum_{i !=j} k\left(x_{\mathrm{s}}(i), x_{\mathrm{s}}(j)\right)-k\left(x_{\mathrm{s}}(i), x_{\mathrm{t}}(j)\right)- \\
k\left(x_{\mathrm{t}}(i), x_{\mathrm{s}}(j)\right)+k\left(x_{\mathrm{t}}(i), x_{\mathrm{t}}(j)\right),$$
where $k$ is a kernel between samples $x_{\mathrm{s}}$ from the source domain and samples $x_{\mathrm{t}}$ from the target domain, and $N$ is the total number of samples across both domains.
We follow~\cite{ciprijanovic2021deepmerge, zhang2020fisher} and use a combination of multiple Gaussian Radial Basis Functions (RBF) kernels.

\section{Data}
\label{sec:data}

\begin{figure}[h]
  \centering
  \includegraphics[scale=0.39]{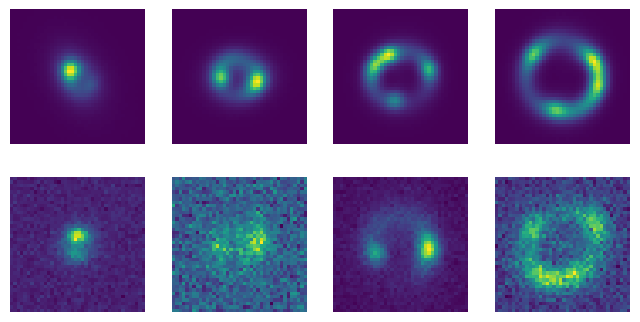}
  \caption{Examples of our source (top) and target (bottom) domain images. The Einstein radius increases from left to right.}
  \label{fig:data}
\end{figure}

We use the {\tt deeplenstronomy}~\cite{morgan2021deeplenstronomy} package, which is built on {\tt lenstronomy}~\cite{birrer2018lenstronomy}, to generate galaxy-scale strong lens datasets with $40\times40$-pixel images in a source domain and in a target domain.
For the source domain, we generate images with no pixel noise.
For the target domain, we add noise based on 
DES\cite{PhysRevD.98.043526} observing conditions, which include sky brightness and seeing, drawn from empirical distributions of these values~\cite{Abbott_2018}.
Both domains contain $50,000$ lenses with Einstein radius varying from 0.5 arc seconds to 3.0 arc seconds drawn from a uniform distribution. 
We divide the dataset into training, validation, and testing with a ratio of 70\%:10\%:20\%. 
Figure~\ref{fig:data} provides example lensing systems from the source and target domain datasets. 

\section{Neural Network Architecture}
\label{sec:NN Architecture}
We employ a baseline CNN with three convolution layers (each layer followed by ReLu activation, batch normalization, and max pooling) and two linear layers. 
Details of the CNN architecture are in Table~\ref{architecture} in the appendix. 
MSE loss is used for the main regression task, and DA is performed using a DANN or MMD loss. 
For both the DANN and MMD variations, latent space features are gathered after the final max pooling layer. 
For the DANN, we use a gradient reversal layer followed by the same two-linear layer architecture repeated for the domain classifier. 
A softmax activation is applied to the domain classifier's output and NLL is used as the domain classifier loss. We use the Adam optimizer for both the DANN and the MMD. Optimal hyperparameters were found through cross-validation. For all models, the batch size is fixed at 32. For training without DA we used a learning rate of $10^{-5}$, while both DANN and MMD used $6\times10^{-5}$.
For MMD, we used DA loss weight $\lambda_{DA}=1.4$. The DA loss weight for DANN was set adaptively as proposed in~\cite{ganin2016domain}. It is initiated at 0 and gradually reaches 1 over the course of training, using the weighting function $\lambda_{DA}=\frac{2}{1+\exp{(-10 \cdot p)}}-1$, where $p$ is fraction of training process completed, from 0 to 1.
For all models, convergence was reached after 30 epochs. 


To evaluate the performance of the CNN, we use the R-squared (R\textsuperscript{2}) score~\cite{wright1921correlation}. It is defined as $R^2=1 - \frac{SS_{Regression}}{SS_{Total}}$, where ${SS_{Regression}=\sum{(\hat{y_i}-y_i)^2}}$ and ${SS_{Total}=\sum{(\bar{y}-y_i)^2}}$. Here, $\hat{y_i}$ is the model's prediction, $y_i$ is the corresponding true value, and $\bar{y}$ is the mean of all true values. This metric shows how well the model can predict a dependent variable (Einstein radius). A larger R\textsuperscript{2} indicates a better fit of the data, with a perfect fit of $R\textsuperscript{2}=1.0$.

\section{Results}
\label{sec:results}

\begin{table}
\centering
\begin{tabular}{lllll}
\hline
                    & \textbf{Source Domain}                            & \textbf{Target Domain}               \\ \hline
\textbf{No DA}          & $0.98\pm 0.0011$                       & $0.50\pm 0.23$                  \\ \hline
\textbf{DANN}          & $0.97\pm 0.0052$                    & {\color[HTML]{333333} \boldmath{$0.94\pm 0.011$}}                 \\ \hline
\textbf{MMD}          & {$0.98\pm 0.013$}                    & {\color[HTML]{333333} \boldmath{$0.95\pm 0.018$}}                 \\ \hline
\\
\end{tabular}
\caption{R\textsuperscript{2} score for training without DA in the top row, with DANN in the middle row, and with MMD in the bottom row. Each score is the mean of ten identical networks, initiated with different random seeds and trained under the same procedure. Standard deviation is also given across these models.}
\label{r2score}
\end{table}

\begin{figure}
    \centering
    \includegraphics[scale=0.42]{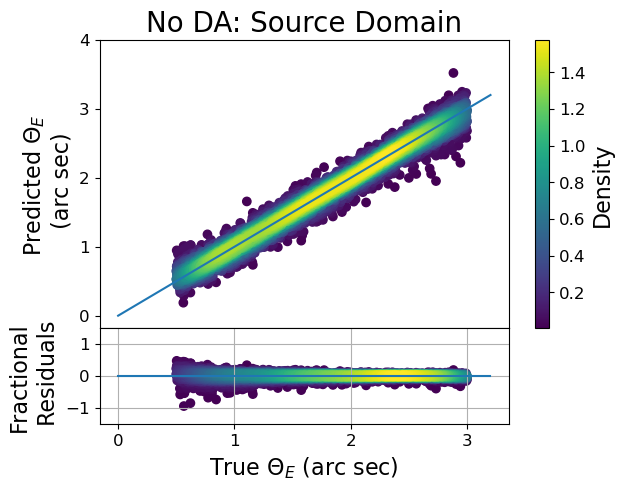}
    \includegraphics[scale=0.42]{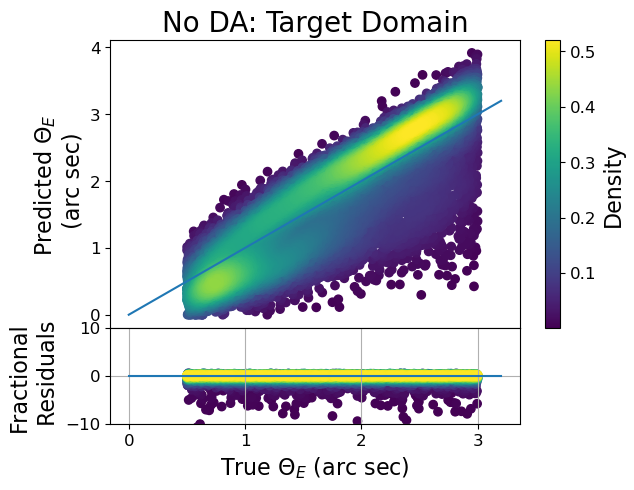}
    \includegraphics[scale=0.42]{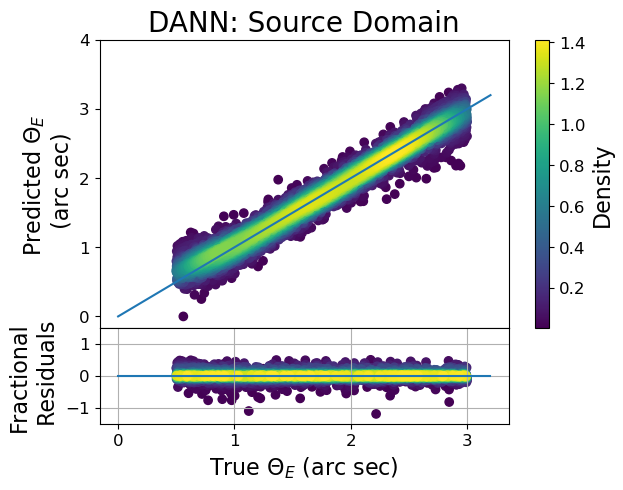}
    \includegraphics[scale=0.42]{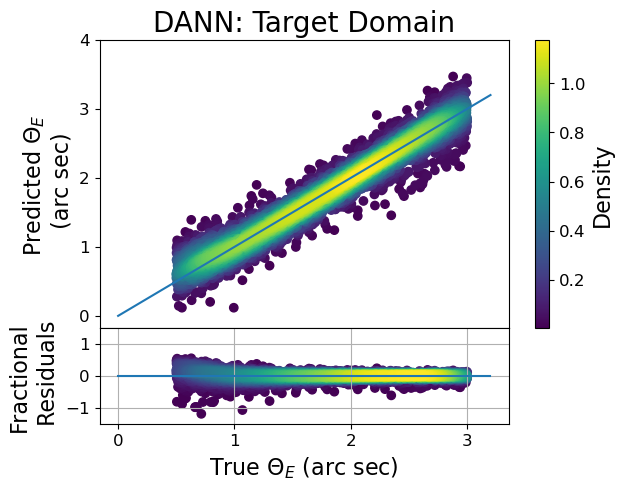}
    \includegraphics[scale=0.42]{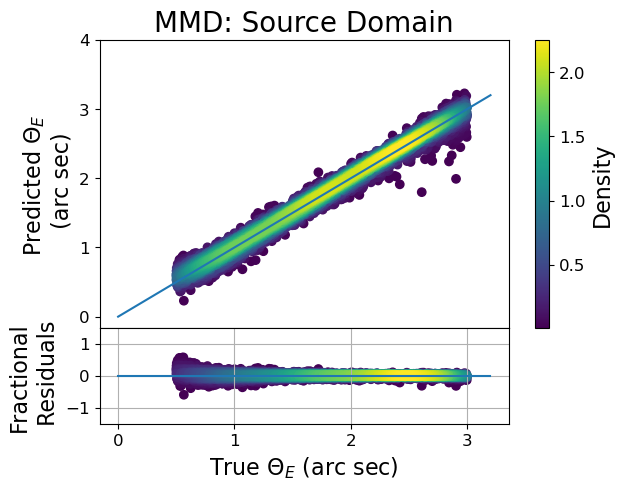}
    \includegraphics[scale=0.42]{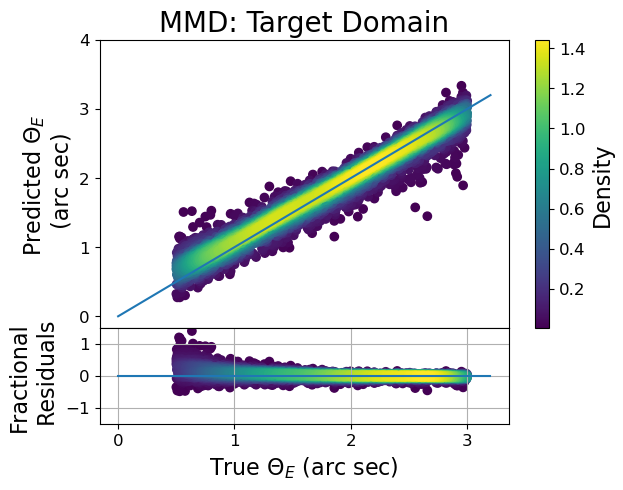}
    \caption{Predictions against actual values of test data sampled from the source domain (left panels) and target domain (right panels). Below each plot are the fractional residuals for visualization. Brighter color represents a higher density of samples, found using a kernel-density estimate with Gaussian kernels~\cite{wkeglarczyk2018kernel}. Results are given without DA (top row), with DANN (middle row), and with MMD (bottom row).}
    \label{fig:predictions}
\end{figure}


We report the R\textsuperscript{2} scores in the source and target test sets in Table~\ref{r2score}. Furthermore, in Figure~\ref{fig:predictions} we visualise predictions against actual values, as well as fractional residuals, which are calculated by $Residual_{fract} = (\Theta_{E,Predicted} - \Theta_{E,True}) / \Theta_{E,Predicted}$. The fractional residuals help visualize how each prediction is potentially biased. 

Figure~\ref{fig:predictions} and the R\textsuperscript{2} scores in Table~\ref{r2score} show that both the DANN and MMD are able to significantly improve model performance on a more complex target domain when trained with a simpler source domain (up to $0.45$ improvement in target $R^2$ scores, making both source and target scores close to $1$) .  
The fractional residuals in Figure~\ref{fig:predictions} support these results, with both DANN and MMD attaining target domain fractional residuals substantially smaller in magnitude than the model without DA. It should also be noted from the residuals that both DANN and MMD struggle the most with lower values of Einstein radius in the target domain. The model without DA and MMD similarly struggle on the low end of Einstein radius in the source domain, while DANN alone performs similarly across the whole source domain. With the inclusion of DA, DANN suffers a small trade-off in accuracy on the source domain. MMD may not lose accuracy on the source data, compared to regular training without DA, but has a higher degree of uncertainty due to its increased standard deviation.


\section{Discussion}
\label{sec:discussion}

These results show that both of the tested DA-based methods have the potential to facilitate the analysis of real survey data in the future -- i.e., reducing bias when using labeled simulated data for training and new unlabeled real data as the test set.

Currently, the existing number of real lens images are likely not enough to use as a target domain for DANN or MMD. However, as future large-scale surveys are released, many more real lens images will become available, with the help of automated identification using deep learning. Studies such as \cite{Davies_2019}, \cite{Petrillo_2018}, and \cite{schaefer2017deep} have demonstrated the viability of these methods. At that point, DANN and MMD may be used for further analysis.

It should be noted that if noise were to be the only difference between simulated and real data, it may be viable to just simulate noise in training data. However, other domain shifts are likely to exist between real and simulated data, which DANN and MMD would be employed to mitigate.
To this end, a further investigation into domain shifts beyond DES-survey noise emulation is needed to ensure the gap from simulated to real data can be closed. 
Both DANN and MMD would likely struggle to overcome the domain shift caused by differences in data distributions of the parameters the network is trying to infer. For instance, if the source and target domain lenses are drawn from different distribution of Einstein radii, the network may have difficulty estimating this parameter correctly. 
This calls for investigation into other domain adaptation techniques, such as instance-based techniques~\cite{de2021adversarial, farahani2021brief}, which will be a topic of our future investigations.

\section*{Acknowledgments and Disclosure of Funding}
This manuscript has been authored by Fermi Research Alliance, LLC under Contract No. DE-AC02-07CH11359 with the U.S. Department of Energy, Office of Science, Office of High Energy Physics. The authors of this paper have committed themselves to performing this work in an equitable, inclusive, and just environment, and we hold ourselves accountable, believing that the best science is contingent on a good research environment.

We acknowledge the Deep Skies Lab as a community of multi-domain experts and collaborators
who have facilitated an environment of open discussion, idea-generation, and collaboration. This
community was important for the development of this project.

Furthermore, we also thank the anonymous referees whose comments helped improve this manuscript.





\medskip
\small
\bibliographystyle{plain}
\bibliography{main}

\begin{thebibliography}{10}

\bibitem{PhysRevD.98.043526}
T.~M.~C. Abbott, F.~B. Abdalla, A.~Alarcon, J.~Aleksi\ifmmode~\acute{c}\else
  \'{c}\fi{}, S.~Allam, S.~Allen, A.~Amara, J.~Annis, J.~Asorey, S.~Avila, and
  et. al.
\newblock Dark energy survey year 1 results: Cosmological constraints from
  galaxy clustering and weak lensing.
\newblock {\em Phys. Rev. D}, 98:043526, Aug 2018.

\bibitem{Abbott_2018}
T.~M.~C. Abbott, F.~B. Abdalla, S.~Allam, A.~Amara, J.~Annis, J.~Asorey,
  S.~Avila, O.~Ballester, M.~Banerji, W.~Barkhouse, and et~al.
\newblock The dark energy survey: Data release 1.
\newblock {\em The Astrophysical Journal Supplement Series}, 239(2):18, nov
  2018.

\bibitem{Aihara_2017}
Hiroaki Aihara, Nobuo Arimoto, Robert Armstrong, St{\'{e} }phane Arnouts,
  Neta~A Bahcall, Steven Bickerton, James Bosch, Kevin Bundy, Peter~L Capak,
  James H~H Chan, and et. al.
\newblock The hyper suprime-cam {SSP} survey: Overview and survey design.
\newblock {\em Publications of the Astronomical Society of Japan}, 70({SP}1),
  sep 2017.

\bibitem{alexander2021domain}
Stephon Alexander, Sergei Gleyzer, Pranath Reddy, Marcos Tidball, and Michael~W
  Toomey.
\newblock Domain adaptation for simulation-based dark matter searches using
  strong gravitational lensing.
\newblock {\em arXiv preprint arXiv:2112.12121}, 2021.

\bibitem{auger2010sloan}
MW~Auger, T~Treu, AS~Bolton, R~Gavazzi, LVE Koopmans, PJ~Marshall,
  LA~Moustakas, and S~Burles.
\newblock The sloan lens acs survey. x. stellar, dynamical, and total mass
  correlations of massive early-type galaxies.
\newblock {\em The Astrophysical Journal}, 724(1):511, 2010.

\bibitem{avestruz2019automated}
Camille Avestruz, Nan Li, Hanjue Zhu, Matthew Lightman, Thomas~E Collett, and
  Wentao Luo.
\newblock Automated lensing learner: automated strong lensing identification
  with a computer vision technique.
\newblock {\em The Astrophysical Journal}, 877(1):58, 2019.

\bibitem{barnabe2011two}
Matteo Barnabe, Oliver Czoske, L{\'e}on~VE Koopmans, Tommaso Treu, and Adam~S
  Bolton.
\newblock Two-dimensional kinematics of slacs lenses--iii. mass structure and
  dynamics of early-type lens galaxies beyond z$\simeq$ 0.1.
\newblock {\em Monthly Notices of the Royal Astronomical Society},
  415(3):2215--2232, 2011.

\bibitem{birrer2018lenstronomy}
Simon Birrer and Adam Amara.
\newblock lenstronomy: Multi-purpose gravitational lens modelling software
  package.
\newblock {\em Physics of the Dark Universe}, 22:189--201, 2018.

\bibitem{Birrer_2022}
Simon Birrer, Suhail Dhawan, and Anowar~J. Shajib.
\newblock The hubble constant from strongly lensed supernovae with
  standardizable magnifications.
\newblock {\em The Astrophysical Journal}, 924(1):2, jan 2022.

\bibitem{ciprijanovic2021deepmerge}
A~{\'C}iprijanovi{\'c}, Diana Kafkes, Kathryn Downey, Sudney Jenkins, Gabriel~N
  Perdue, Sandeep Madireddy, Travis Johnston, Gregory~F Snyder, and Brian Nord.
\newblock Deepmerge--ii. building robust deep learning algorithms for merging
  galaxy identification across domains.
\newblock {\em Monthly Notices of the Royal Astronomical Society},
  506(1):677--691, 2021.

\bibitem{CL2023}
A.~{{\'C}iprijanovi{\'c}}, A.~{Lewis}, K.~{Pedro}, S.~{Madireddy}, B.~{Nord},
  G.~N. {Perdue}, and S.~M. {Wild}.
\newblock {DeepAstroUDA: semi-supervised universal domain adaptation for
  cross-survey galaxy morphology classification and anomaly detection}.
\newblock {\em Machine Learning: Science and Technology}, 4(2):025013, June
  2023.

\bibitem{CK2022}
Aleksandra {{\'C}iprijanovi{\'c}}, Diana {Kafkes}, Gregory {Snyder}, F.~Javier
  {S{\'a}nchez}, Gabriel~Nathan {Perdue}, Kevin {Pedro}, Brian {Nord}, Sandeep
  {Madireddy}, and Stefan~M. {Wild}.
\newblock {DeepAdversaries: examining the robustness of deep learning models
  for galaxy morphology classification}.
\newblock {\em Machine Learning: Science and Technology}, 3(3):035007,
  September 2022.

\bibitem{Collett_2015}
Thomas~E. Collett.
\newblock {THE} {POPULATION} {OF} {GALAXY}{\textendash}{GALAXY} {STRONG}
  {LENSES} {IN} {FORTHCOMING} {OPTICAL} {IMAGING} {SURVEYS}.
\newblock {\em The Astrophysical Journal}, 811(1):20, sep 2015.

\bibitem{collett2014cosmological}
Thomas~E Collett and Matthew~W Auger.
\newblock Cosmological constraints from the double source plane lens sdssj0946+
  1006.
\newblock {\em Monthly Notices of the Royal Astronomical Society},
  443(2):969--976, 2014.

\bibitem{CS2017}
Gabriela {Csurka}.
\newblock {Domain Adaptation for Visual Applications: A Comprehensive Survey}.
\newblock {\em arXiv e-prints}, page arXiv:1702.05374, February 2017.

\bibitem{Davies_2019}
Andrew Davies, Stephen Serjeant, and Jane~M Bromley.
\newblock Using convolutional neural networks to identify gravitational lenses
  in astronomical images.
\newblock {\em Monthly Notices of the Royal Astronomical Society},
  487(4):5263–5271, May 2019.

\bibitem{de2021adversarial}
Antoine de~Mathelin, Guillaume Richard, Fran{\c{c}}ois Deheeger, Mathilde
  Mougeot, and Nicolas Vayatis.
\newblock Adversarial weighting for domain adaptation in regression.
\newblock In {\em 2021 IEEE 33rd International Conference on Tools with
  Artificial Intelligence (ICTAI)}, pages 49--56. IEEE, 2021.

\bibitem{farahani2021brief}
Abolfazl Farahani, Sahar Voghoei, Khaled Rasheed, and Hamid~R Arabnia.
\newblock A brief review of domain adaptation.
\newblock {\em Advances in data science and information engineering:
  proceedings from ICDATA 2020 and IKE 2020}, pages 877--894, 2021.

\bibitem{ganin2016domain}
Yaroslav Ganin, Evgeniya Ustinova, Hana Ajakan, Pascal Germain, Hugo
  Larochelle, Fran{\c{c}}ois Laviolette, Mario Marchand, and Victor Lempitsky.
\newblock Domain-adversarial training of neural networks.
\newblock {\em The journal of machine learning research}, 17(1):2096--2030,
  2016.

\bibitem{GD2021}
Sankalp {Gilda}, Antoine {de Mathelin}, Sabine {Bellstedt}, and Guillaume
  {Richard}.
\newblock {Unsupervised Domain Adaptation for Constraining Star Formation
  Histories}.
\newblock {\em arXiv e-prints}, page arXiv:2112.14072, December 2021.

\bibitem{JMLR:v13:gretton12a}
Arthur Gretton, Karsten~M. Borgwardt, Malte~J. Rasch, Bernhard Sch{{\"o}}lkopf,
  and Alexander Smola.
\newblock A kernel two-sample test.
\newblock {\em Journal of Machine Learning Research}, 13(25):723--773, 2012.

\bibitem{ivezic2019lsst}
{\v{Z}}eljko Ivezi{\'c}, Steven~M Kahn, J~Anthony Tyson, Bob Abel, Emily
  Acosta, Robyn Allsman, David Alonso, Yusra AlSayyad, Scott~F Anderson, John
  Andrew, et~al.
\newblock Lsst: from science drivers to reference design and anticipated data
  products.
\newblock {\em The Astrophysical Journal}, 873(2):111, 2019.

\bibitem{Lin_2021}
Qiufan Lin, Dominique Fouchez, and Jerome Pasquet.
\newblock Galaxy image translation with semi-supervised noise-reconstructed
  generative adversarial networks.
\newblock In {\em 2020 25th International Conference on Pattern Recognition
  (ICPR)}. IEEE, January 2021.

\bibitem{morgan2021deeplenstronomy}
Robert Morgan, Brian Nord, Simon Birrer, Joshua Yao-Yu Lin, and Jason Poh.
\newblock deeplenstronomy: A dataset simulation package for strong
  gravitational lensing.
\newblock {\em arXiv preprint arXiv:2102.02830}, 2021.

\bibitem{O_Briain_2021}
Teaghan O’Briain, Yuan-Sen Ting, Sébastien Fabbro, Kwang~M. Yi, Kim Venn,
  and Spencer Bialek.
\newblock Cycle-starnet: Bridging the gap between theory and data by leveraging
  large data sets.
\newblock {\em The Astrophysical Journal}, 906(2):130, January 2021.

\bibitem{2019clrp.2020...20P}
Will {Percival}, Michael {Balogh}, Dick {Bond}, Jo~{Bovy}, Raymond {Carlberg},
  Scott {Chapman}, Patrick {Cote}, Nicolas {Cowan}, Sebastien {Fabbro}, Laura
  {Ferrarese}, and {et al.}
\newblock {The Euclid Mission}.
\newblock In {\em Canadian Long Range Plan for Astronomy and Astrophysics White
  Papers}, volume 2020, page~20, October 2019.

\bibitem{Petrillo_2018}
C~E Petrillo, C~Tortora, S~Chatterjee, G~Vernardos, L~V~E Koopmans, G~Verdoes
  Kleijn, N~R Napolitano, G~Covone, L~S Kelvin, and A~M Hopkins.
\newblock Testing convolutional neural networks for finding strong
  gravitational lenses in kids.
\newblock {\em Monthly Notices of the Royal Astronomical Society}, October
  2018.

\bibitem{Rezaei_2022}
S~Rezaei, J~P McKean, M~Biehl, W~de~Roo, and A~Lafontaine.
\newblock A machine learning based approach to gravitational lens
  identification with the international {LOFAR} telescope.
\newblock {\em Monthly Notices of the Royal Astronomical Society},
  517(1):1156--1170, jul 2022.

\bibitem{schaefer2017deep}
C~Schaefer, M~Geiger, T~Kuntzer, and JP~Kneib.
\newblock Deep convolutional neural networks as strong gravitational lens
  detectors.
\newblock {\em arXiv preprint arXiv:1705.07132}, 2017.

\bibitem{SN2023}
Manuel {Schwonberg}, Joshua {Niemeijer}, Jan-Aike {Term{\"o}hlen}, J{\"o}rg~P.
  {Sch{\"a}fer}, Nico~M. {Schmidt}, Hanno {Gottschalk}, and Tim {Fingscheidt}.
\newblock {Survey on Unsupervised Domain Adaptation for Semantic Segmentation
  for Visual Perception in Automated Driving}.
\newblock {\em arXiv e-prints}, page arXiv:2304.11928, April 2023.

\bibitem{sereno2002probing}
Mauro Sereno.
\newblock Probing the dark energy with strong lensing by clusters of galaxies.
\newblock {\em arXiv preprint astro-ph/0209210}, 2002.

\bibitem{suyu2010dissecting}
SH~Suyu, PJ~Marshall, MW~Auger, S~Hilbert, RD~Blandford, LVE Koopmans,
  CD~Fassnacht, and T~Treu.
\newblock Dissecting the gravitational lens b1608+ 656. ii. precision
  measurements of the hubble constant, spatial curvature, and the dark energy
  equation of state.
\newblock {\em The Astrophysical Journal}, 711(1):201, 2010.

\bibitem{Vilalta_2019}
Ricardo Vilalta, Kinjal~Dhar Gupta, Dainis Boumber, and Mikhail~M. Meskhi.
\newblock A general approach to domain adaptation with applications in
  astronomy.
\newblock {\em Publications of the Astronomical Society of the Pacific},
  131(1004):108008, sep 2019.

\bibitem{wkeglarczyk2018kernel}
Stanis{\l}aw W{\k{e}}glarczyk.
\newblock Kernel density estimation and its application.
\newblock In {\em ITM web of conferences}, volume~23, page 00037. EDP Sciences,
  2018.

\bibitem{Wilde_2022}
Joshua Wilde, Stephen Serjeant, Jane~M Bromley, Hugh Dickinson, L{\'{e} }on V~E
  Koopmans, and R~Benton Metcalf.
\newblock Detecting gravitational lenses using machine learning: exploring
  interpretability and sensitivity to rare lensing configurations.
\newblock {\em Monthly Notices of the Royal Astronomical Society},
  512(3):3464--3479, feb 2022.

\bibitem{wright1921correlation}
S~Wright.
\newblock Correlation and causation.
\newblock {\em Journal of agricultural research}, 20(7):557, 1921.

\bibitem{zaborowski2022identification}
E~Zaborowski, A~Drlica-Wagner, F~Ashmead, JF~Wu, R~Morgan, CR~Bom, AJ~Shajib,
  S~Birrer, W~Cerny, L~Buckley-Geer, et~al.
\newblock Identification of galaxy-galaxy strong lens candidates in the decam
  local volume exploration survey using machine learning.
\newblock {\em arXiv preprint arXiv:2210.10802}, 2022.

\bibitem{zhang2020fisher}
Yinghua Zhang, Yu~Zhang, Ying Wei, Kun Bai, Yangqiu Song, and Qiang Yang.
\newblock Fisher deep domain adaptation, 2020.

\end{thebibliography}
\newpage
\section*{Appendix}

\begin{table}[h]
\centering
\begin{tabular}{llllll}
\hline
\textbf{Layers}     & \textbf{Properties}                                                              & \textbf{Output Shape} & \textbf{Parameters} \\ \hline
input               & 1*40*40                                                                          & (1, 40, 40)           & 0                   \\ \hline
Convolution (2D)    & \begin{tabular}[c]{@{}l@{}}Filters: 8\\ Kernel: 3 3\\ Activation: ReLu\end{tabular}                 & (8, 42, 42)           & 80                  \\ \hline
Batch Normalization & -                                                                                                  & (8, 42, 42)           & 16                  \\ \hline
MaxPooling          & \begin{tabular}[c]{@{}l@{}}Kernel: 2 2\\ Stride: 2\end{tabular}                                                                                & (8, 21, 21)           & 0                   \\ \hline
Convolution (2D)    & \begin{tabular}[c]{@{}l@{}}Filters: 16\\ Kernel: 3 3\\ Activation: ReLu\end{tabular}                & (16, 21, 21)          & 1,168               \\ \hline
Batch Normalization & -                                                                                             & (16, 21, 21)          & 32                  \\ \hline
MaxPooling          & \begin{tabular}[c]{@{}l@{}}Kernel: 2 2\\ Stride: 2\end{tabular}                                                                                      & (16, 10, 10)          & 0                   \\ \hline
Convolution (2D)    & \begin{tabular}[c]{@{}l@{}}Filters: 32\\ Kernel: 3 3\\ Activation: ReLu\end{tabular}             & (32, 10, 10)          & 4,640               \\ \hline
Batch Normalization & -                                                                       & (32, 10, 10)          & 64                  \\ \hline
MaxPooling          & \begin{tabular}[c]{@{}l@{}}Kernel: 2 2\\ Stride: 2\end{tabular}                                                                        & (32, 5, 5)            & 0                   \\ \hline
Flatten             & -                                                                               & (800)                 & -                   \\ \hline
Fully Connected     & Activation: ReLu                                                                  & (128)                 & 102,528              \\ \hline
Fully Connected     & Activation: ReLu                                                     & (1)                   & 129                 \\ \hline
\end{tabular}
\caption{CNN architecture}
\label{architecture}
\end{table}

\end{document}